\documentclass[12pt]{iopart}

\usepackage{graphicx}
\begin{document}

\newcommand{\eli}{$\acute{{\rm E}}$liashberg }
\renewcommand{\k}{\vec{k}}
\newcommand{\kk}{\vec{k'}}
\newcommand{\q}{\vec{q}}
\newcommand{\Q}{\vec{Q}}
\renewcommand{\r}{\vec{r}}
\newcommand{\s}{{\mit{\it \Sigma}}}
\newcommand{\Tc}{$T_{\rm c}$ }
\newcommand{\Tcf}{$T_{\rm c}$}
\newcommand{\Science}{{\it Science} } 
\newcommand{\Nature}{{\it Nature} } 
\newcommand{\qf}{\vec{q}_{\rm FFLO}}
\newcommand{\qi}{\vec{q}_{\rm inc}}
\newcommand{\qia}{\vec{q}_{\rm inc}^{\rm \,\,a}}
\newcommand{\qib}{\vec{q}_{\rm inc}^{\rm \,\,b}}
\newcommand{\Hc}{$H_{\rm c2}^{\rm P}$ }
\newcommand{\Hcf}{$H_{\rm c2}^{\rm P}$}
\newcommand{\Co}{CeCoIn$_5$ }
\newcommand{\Cof}{CeCoIn$_5$}
\newcommand{\va}{\vec{a}}
\newcommand{\vb}{\vec{b}}
\renewcommand{\i}{\hspace*{0.3mm}\vec{i}\hspace*{0.6mm}}
\renewcommand{\j}{\hspace*{0.3mm}\vec{j}\hspace*{0.6mm}}

\title[]{Magnetic structure of antiferromagnetic 
Fulde-Ferrell-Larkin-Ovchinnikov state}

\author{Youichi Yanase$^{1,2}$ and Manfred Sigrist$^2$}

\address{$^1$ Department of Physics, Niigata University, 
         Niigata 950-2041, Japan}
\address{$^2$ Theoretische Physik, ETH Zurich, 8093 Zurich,
         Switzerland}

\ead{yanase@phys.sc.niigata-u.ac.jp}

\begin{abstract}
The properties of incommensurate antiferromagnetic (AFM) order in the 
Fulde-Ferrell-Larkin-Ovchinnikov (FFLO) state is studied by solving
the Bogoliubov-de-Gennes (BdG) equations. 
The relationship between the electronic structure 
and the magnetic structure is clarified. 
We find that the magnetic structure in the AFM-FFLO state 
includes three cases. 
(I) In the strongly localized case, the AFM staggered moment is 
confined into the FFLO nodal planes where the superconducting order 
parameter vanishes. 
(II) In the weakly localized case, the AFM staggered moment appears 
in the whole spatial region, and its magnitude is enhanced around the 
FFLO nodal planes. 
(III) In the extended case, the AFM staggered moment is nearly 
homogeneous and slightly suppressed in the vicinity of FFLO nodal 
planes. 
 The structure of Bragg peaks in the momentum resolved structure 
factor is studied in each case. 
 We discuss the possibility of AFM-FFLO state in the heavy fermion  
superconductor CeCoIn$_5$ by comparing these results with 
the neutron scattering data of CeCoIn$_5$. 
Experimentally the magnetic structure and its dependence on the magnetic field orientation in the high field superconducting phase of CeCoIn$_5$ are 
consistent with the case (II).

\end{abstract}


\section{Introduction}

The possible presence of a spatially modulated state in superconductors in a high magnetic field
was predicted by Fulde and Ferrell~\cite{FF}, and by 
Larkin and Ovchinnikov~\cite{LO} more than 40 years ago. 
 While the Bardeen-Cooper-Schrieffer (BCS) theory assume Cooper pairs with vanishing 
 total momentum, the FFLO superconducting state represents a
condensate of Cooper pairs with a finite total momentum. 
 Since the FFLO state has an internal degree of freedom arising from the
reflection or inversion symmetry, a spontaneous breaking of 
the spatial symmetry occurs.  
 Although this novel superconducting state with an exotic symmetry has 
been attracting much interest, the FFLO state has not been observed in
superconductors for nearly 40 years. 
 Under these circumstances, the discovery of a new superconducting phase 
in \Co at high magnetic fields and low temperatures
~\cite{radovan2003,PhysRevLett.91.187004} triggered many theoretical 
and experimental studies because this high-field superconducting (HFSC) 
phase is a likely candidate for the FFLO state~\cite{matsuda2007}. 
 The recent interest on the FFLO superconductivity/superfluidity  
extends further into various related fields, such as organic 
superconductors~\cite{uji:157001,singleton2000,lortz:187002,
shinagawa:147002,yonezawa:117002}, 
cold atom gases~\cite{Partridge01272006,Zwierlein01272006}, 
astrophysics, and nuclear physics \cite{casalbuoni2004}.

 The HFSC phase of \Co has been interpreted widely within the concept of 
the FFLO state~\cite{matsuda2007,watanabe2004,capan2004,martin2005,
mitrovic2006,miclea2006,correa2007,mitrovic2008,adachi2003,
ikeda:134504,ikeda:054517}. 
 However, recent observations of the magnetic order in 
the HFSC phase call for a reexamination of this conclusion
~\cite{young2007,kenzelmann2008}. 
It is not unlikely that this order will be closely connected with a
AFM quantum critical point observed 
in \Cof ~\cite{bianchi2003,ronning2005}. 
Moreover, the nuclear magnetic resonance 
(NMR)~\cite{young2007,koutroulakis2010,kumagaiprivate} 
and neutron scattering~\cite{kenzelmann2008,kenzelmann2010} 
measurements may have uncovered a novel superconducting state in 
this strongly correlated electron system.

Neutron scattering measurements have found that 
the wave vector of the AFM order is incommensurate 
$\q_{\rm AF} = \Q + \qi$ with $\Q = (\pi,\pi)$ and the AFM staggered 
moment $\vec{M}_{\rm AF}$ is oriented along 
the {\it c}-axis~\cite{kenzelmann2008}. 
Recent experiments have shown that the 
incommensurability $\qi$ is fixed along [1,-1,0] irrespective whether magnetic 
field is directed along [1,1,0] and [1,0,0] in the tetragonal 
lattice~\cite{kenzelmann2010}.
 These magnetic structures are consistent with 
the NMR measurements~\cite{curro2009}.

 Some theoretical scenarios have been proposed to explain the AFM order in the HFSC phase 
of \Co. We have analyzed the possibility that AFM order arises from the inhomogeneous 
Larkin-Ovchinnikov (LO) state~\cite{yanase_LT,yanase_JPSJ2009}. 
 The AFM order triggered by the emergence of $\pi$-triplet pairing or 
pair density wave (PDW) has been investigated in the BCS 
state~\cite{aperis2008,aperis2010,agterberg2009} and 
in the homogeneous Fulde-Ferrell (FF) state~\cite{miyake2008}.  
 In order to identify the HFSC phase of \Co it is highly desirable 
to examine these possible phases by comparing their properties with 
the experimental results. 
 In this study, we investigate the magnetic structure of AFM-FFLO state, 
in which the AFM order appears in the inhomogeneous LO state, 
and discuss the recent neutron scattering measurements.

\section{Formulation}

 Our theoretical analysis is based on the microscopic model,
\begin{eqnarray}
  \label{eq:model}
  && \hspace{-15mm}
  H= - t \sum_{\langle \i,\j \rangle,\sigma} c_{\i,\sigma}^{\dag}c_{\j,\sigma}
  + t' \sum_{\langle\langle \i,\j \rangle\rangle,\sigma} c_{\i,\sigma}^{\dag}c_{\j,\sigma} 
  + t'' \sum_{\langle\langle\langle \i,\j \rangle\rangle\rangle,\sigma} 
c_{\i,\sigma}^{\dag}c_{\j,\sigma} 
 - \mu \sum_{\i,\sigma} c_{\i,\sigma}^{\dag}c_{\i,\sigma}
\nonumber \\ && \hspace{-10mm} 
  + U \sum_{\i} n_{{\i}\uparrow} n_{{\i}\downarrow} 
  + V \sum_{\langle \i,\j \rangle} n_{\i} \hspace{0.5mm} n_{\j} 
  + J \sum_{\langle \i,\j \rangle} \vec{S}_{\i} \hspace{0.4mm} \vec{S}_{\j} 
  - g_{\rm B} \vec{H} \sum_{\i} \vec{S}_{\i}, 
\end{eqnarray}
where $\vec{S}_{\i}$ 
is the spin operator and $n_{\i}$ is the number operator 
at site $\i = (m,n)$. 
 To describe the quasi-two-dimensional electronic structure 
of CeCoIn$_5$, we assume a square lattice, in which 
the bracket $\langle \i,\j \rangle$, $\langle\langle \i,\j \rangle\rangle$, 
and $\langle\langle\langle \i,\j \rangle\rangle\rangle$
denote the summation over the 
nearest-neighbor sites, next-nearest-neighbor sites, and third nearest 
neighbor sites, respectively. 

 The on-site repulsive interaction is given by $U$, and 
$V$ and $J$ stand for the attractive and AFM  exchange interactions, 
respectively, between nearest-neighbor sites. 
 We introduce $V$ to stabilize the $d$-wave superconducting state within the 
mean field BdG equations and choose $J > 0$  
favoring AFM correlation in CeCoIn$_5$. 
 The effective interaction leading to the $d$-wave superconductivity 
near the AFM instability arises from the simple Hubbard model or 
periodical Anderson model beyond the mean field theory
~\cite{yanase2003cms,yanaseFFLOQCP}, but we here assume 
the interactions $V$ and $J$ to describe these features 
in the inhomogeneous LO phase on the basis of mean field BdG equations. 
 Although the BdG equations neglect the AFM spin fluctuation 
beyond the mean field approximation, 
they are suitable for studying the qualitative features of the 
inhomogeneous superconducting and/or magnetic state. 
 The roles of AFM spin fluctuation has been discussed 
in Ref.~\cite{yanase_JPSJ2009}.

With the last term in eq.~(1), we include the Zeeman coupling due to the 
applied magnetic field. The $g$-factor is assumed to be $g_{\rm B}=2$. 
The magnetic field lies in the {\it ab}-plane of 
the tetragonal lattice and  the superconducting vortices are neglected
for simplicity. 
 We choose the unit of energy such that $t=1$ and $t'/t=0.25$, $t''/t=0.05$. 
 The chemical potential enters as $\mu=\mu_{0} + (\frac{1}{2} U + 4 V) n_{0}$, 
where $n_{0}$ is the number density for $U=V=J=H=0$. 
 We vary the bare chemical potential from $\mu_{0} = -1.25$ to 
$\mu_{0} = -1.15$ to investigate the possible magnetic structures 
in the AFM-FFLO state. 
 Then, we obtain the number density $n = 0.72 \sim 0.82$.

We study the magnetic structure in the AFM-FFLO state in the following way. 
First, the BdG equations are self-consistently solved 
for the mean fields of the spin $\langle S_{\i}^{\rm h} \rangle$, 
charge $\langle n_{\i} \rangle$, and superconductivity 
$\Delta_{\i,\j}^{\sigma\sigma'} =
\langle c_{{\i}\sigma}c_{{\j}\sigma'} \rangle$, 
where $S_{\i}^{\rm h}$ is the spin operator parallel to the magnetic field. 
 We take into account the Hartree term arising from $U$, $V$, and $J$. 
 Since the spin singlet $d$-wave state is stable in this model, 
the superconducting order parameter is described as
\begin{eqnarray}
  \label{eq:dSC}
  && \hspace{0mm}
\Delta^{\rm d}(\i) 
= \Delta_{\i,\i+\va}^{\uparrow\downarrow} 
+ \Delta_{\i,\i-\va}^{\uparrow\downarrow} 
- \Delta_{\i,\i+\vb}^{\uparrow\downarrow} 
- \Delta_{\i,\i-\vb}^{\uparrow\downarrow},  
\end{eqnarray}
with $\va$ and $\vb$ being the unit vector along the {\it a}- 
and {\it b}-axes, respectively. 
 As for the spatial dependence of $\Delta^{\rm d}(\i)$, 
we find that the inhomogeneous LO state is stable against the 
uniform FF state. Then, the order parameter is real and oscillates 
in the space. 
 We assume the single-$q$ spatial modulation along the magnetic field 
in which the order parameter is approximately described as 
$\Delta^{\rm d}(\i) \sim \Delta \cos({\rm i}\qf \cdot \i)$ 
with $\qf \parallel \vec{H}$ except for the vicinity of 
the BCS-FFLO transition. 
 This is the most stable FFLO state in the presence of 
superconducting vortices when the Fermi surface is nearly isotropic. 
 We determine the stable superconducting state by comparing the 
condensation energy of BCS, FFLO and normal states. 
 The orientation of magnetic field is taken into account 
in the direction of $\qf$ in our formulation. 
 The case of $\qf$ not parallel to $\vec{H}$ will be discussed 
at the last of this paper.

 Second, we determine the magnetic instability by calculating the 
transverse spin susceptibility using the random phase 
approximation,
\begin{eqnarray} 
&& \hspace{0mm}
\chi^{\pm}(\i,\j) = 
\frac{\chi^{\pm}_{0}(\i,\j)}{1 - \sum_{\k} I(\i,\k) \chi^{\pm}_{0}(\k,\j)}
\end{eqnarray}
where $I(\i,\i) = U$, $I(\i,\i \pm \va) = I(\i,\i \pm \vb) = -\frac{J}{2}$, 
and otherwise $I(\i,\k) = 0$. 
 The bare spin susceptibility $\chi^{\pm}_{0}(\k,\j)$ is calculated for 
the mean field Hamiltonian obtained by solving the BdG equations. 
 The magnetic instability is determined by the divergence of transverse 
spin susceptibility assuming a second order magnetic phase transition. 
 Then, the criterion is $\lambda_{\rm max} =1$ 
where $\lambda_{\rm max}$ is the maximum eigenvalue of the matrix 
$K(\i,\j) = \sum_{\k} I(\i,\k) \chi^{\pm}_{0}(\k,\j)$. 
 The magnetic moment perpendicular to the magnetic field is obtained 
as 
\begin{eqnarray}
 \label{eq:AFM}
&& \hspace{0mm}
M(\i) = I^{-1}(\i,\k) m(\k), 
\end{eqnarray}
near the critical point, 
where $m(\k)$ is the eigenvector of $K(\i,\j)$ for the eigenvalue 
$\lambda_{\rm max}$ and $\sum_{\k} I^{-1}(\i,\k) I(\k,\j) = \delta_{\i,\j}$. 
 We denote the AFM staggered moment $M_{\rm AF}(\i)  = (-1)^{m+n} M(\i)$ 
with $\i = (m,n)$.

 Since the effect of induced PDW order, which plays an important role 
to stabilize the AFM-FFLO state, is ignored in this formulation, 
the stability of the AFM order is underestimated. 
 However, we confirmed that the induced PDW does not affect the 
magnetic structure~\cite{yanase_JPSJ2009}.  
 Choosing the parameters $U$, $V$, and $J$ so that 
$\lambda_{\rm max} =1$ in the FFLO state as in Refs.~\cite{yanase_LT,yanase_JPSJ2009},  
we discuss the variety of magnetic structures in the AFM-FFLO state 
in the following sections. 
 In our model the incommensurate wave vector $\qi$ is along $\va$ or $\vb$ 
direction and different from the experimental observation 
$\qi \parallel [1,\pm 1,0]$~\cite{kenzelmann2008}. 
 However, the relationship between the magnetic structure and 
relative angle of $\qi$ and $\qf$ (and $\vec{H}$) is appropriately 
captured in our calculation.

\section{Magnetic structure}

 We here show that the magnetic structure in the AFM-FFLO state 
is classified into three cases. 
 (I) In the strongly localized case, the AFM moment is 
confined into the FFLO nodal planes where the superconducting order 
parameter vanishes. 
 The magnitude of AFM moment away from the nodal planes 
is typically $\sim 0.01$ of that around the nodal planes. 
 (II) In the weakly localized case, the AFM moment appears 
in the whole spatial region, and its magnitude is enhanced around the 
FFLO nodal planes, typically twice. 
 (III) In the extended case, the AFM moment is nearly 
homogeneous and slightly suppressed in the vicinity of FFLO nodal 
planes.

\begin{figure}[ht]
\begin{center}
\includegraphics[width=16cm]{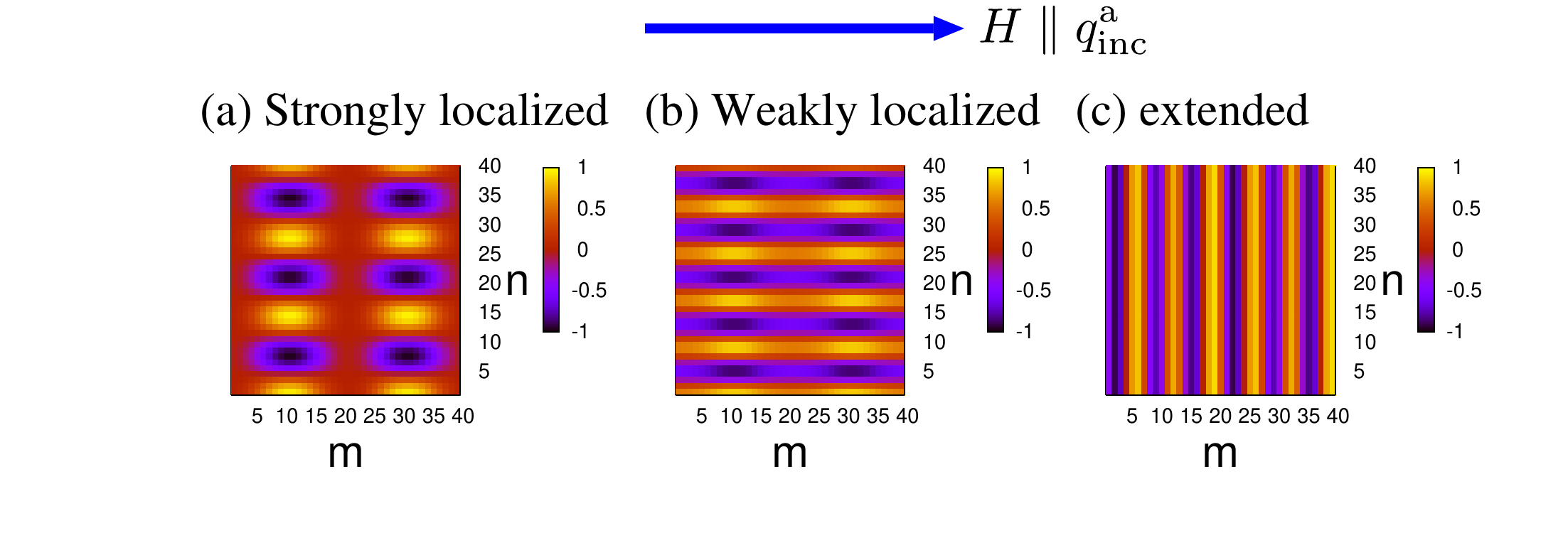}\hspace{1pc}%
\caption{
The AFM staggered moment normalized by its maximum value 
$M_{\rm AF}(\i)/M_{\rm max}$ with $\i = (m,n)$ 
in the case of $\qf \parallel \qia$. 
 This FFLO modulation corresponds to the experimental setup of \Co 
for the magnetic field along $[1,\pm 1,0]$ direction~\cite{kenzelmann2008}.  
(a), (b), and (c) show (I) strongly localized case, 
(II) weakly localized case, and (III) extended case, respectively. 
We solve the BdG equations for $40 \time 40$ lattices. 
We choose the parameters 
(a) $\mu_0 = -1.15$, $U=0.9$, $J=0.6$, and $V=-0.5$,  
(b) $\mu_0 = -1.19$, $U=1.05$, $J=0.6$, and $V=-0.5$,  
and (c) $\mu_0 = -1.25$, $U=1.15$, $J=0.65$, and $V=-0.5$
so that the AFM order occurs in the FFLO state at 
(a) $(T,H)=(0.0315,0.164)$, (b) $(T,H)=(0.025,0.155)$, and 
(c) $(T,H)=(0.02,0.168)$, respectively. 
The arrow shows the direction of magnetic field. 
}
\end{center}
\end{figure}

  Owing to the four fold rotation symmetry of the tetragonal lattice 
the incommensurate wave vector $\qi = \qia \parallel \va$ is degenerate with 
$\qi = \qib \parallel \vb$ in the normal state and uniform BCS state. 
 This degeneracy is lifted in the FFLO state except for 
$\qf \parallel \qia + \qib$. 
 Figures 1(a), 1(b), and 1(c) show typical magnetic structures of 
cases (I), (II), and (III), respectively 
for the magnetic field $\vec{H} \parallel \qia$. 
 We see that the modulation vector of AFM moment is 
different between the case (III) and the other two cases. 
 In the extended case the incommensurate wave vector $\qi$ is parallel to 
the FFLO modulation vector $\qf$. 
 Thus, the extended case (III) is incompatible with 
the neutron scattering measurement~\cite{kenzelmann2008}, 
assuming the relation $\vec{H} \parallel \qf$. 
 On the other hand, the incommensurate wave vector $\qi$ is perpendicular 
to the FFLO modulation vector $\qf$ in the cases (I) and (II),
consistent with the neutron scattering measurement for 
$\vec{H} \parallel [1,\pm 1,0]$. 
 This means that the neutron scattering measurement 
indicates the localized AFM moment around the FFLO nodal planes. 
 
\begin{figure}[ht]
\begin{center}
\includegraphics[width=16cm]{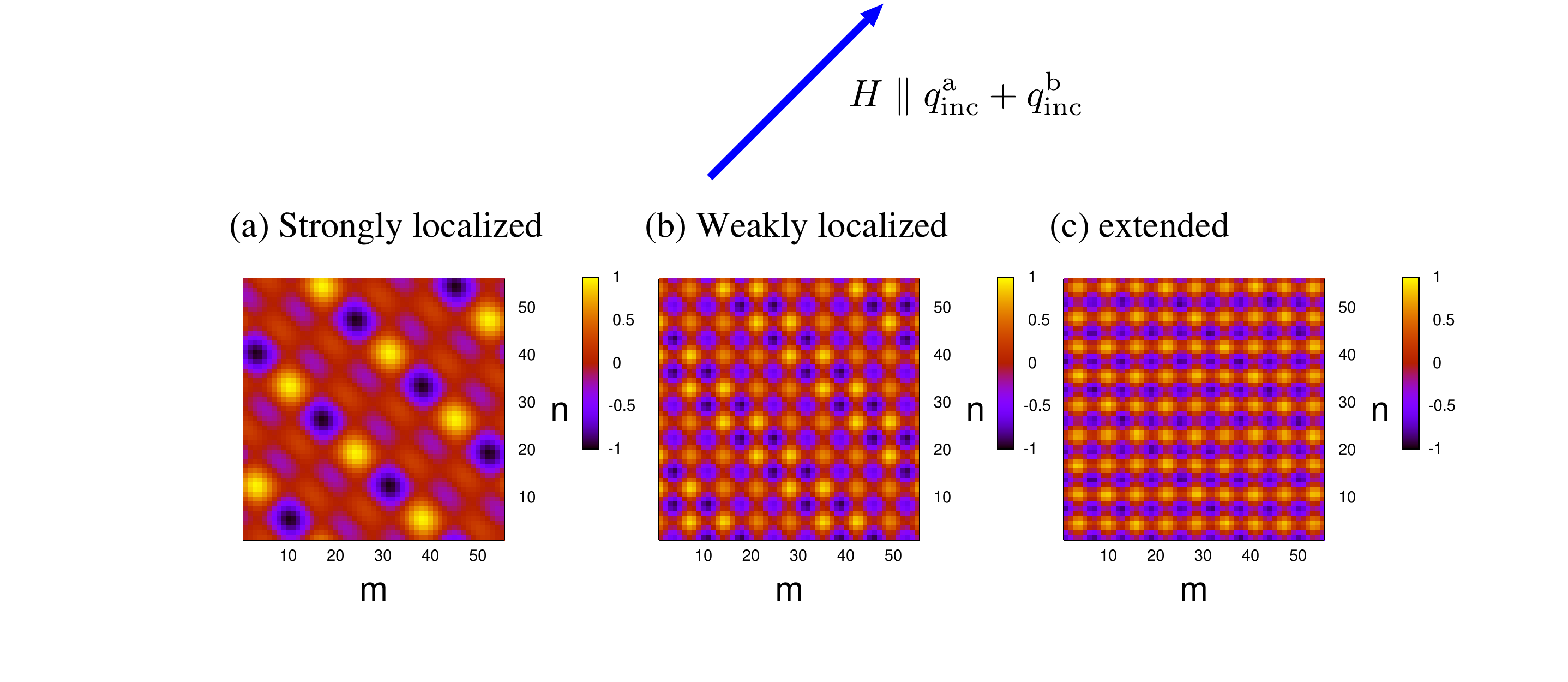}\hspace{1pc}%
\caption{
The normalized AFM staggered moment $M_{\rm AF}(\i)/M_{\rm max}$ 
in the case of $\qf \parallel \qia+\qib$. 
This FFLO modulation corresponds to the \Co 
in the magnetic field along $[1,0,0]$ or $[0,1,0]$ direction 
~\cite{young2007,kenzelmann2010}.  
We assume the same parameter as in Fig.~1. 
The AFM order occurs at (a) $(T,H)=(0.0264,0.164)$, 
(b) $(T,H)=(0.0324,0.155)$, and 
(c) $(T,H)=(0.035,0.15)$, respectively. 
We solve the BdG equations for $56 \times 56$ lattices 
to keep the amplitude of FFLO modulation vector $|\qf|$ similar to in Fig.~1. 
The arrow shows the direction of magnetic field. 
}
\end{center}
\end{figure}
 
 Figure 2 shows the magnetic structure for the magnetic field 
$\vec{H} \parallel \qia + \qib$. 
 We see that the AFM staggered moment $M_{\rm AF}(\i)$ is distributed 
similarly to Fig.~1. 
 The magnetic moment is strongly (weakly) localized in Fig.~2(a) 
(Fig.~2(b)), while that is extended in Fig.~2(c).
 Figs.~2(b) and 2(c) show double-$q$ structure like 
$M_{\rm AF}(\i) \sim \cos(\qia \cdot \i) \pm \cos(\qib \cdot \i)$ 
in contrast to Figs.~1(b) and 1(c). 
 This double-$q$ structure arises from the translational symmetry breaking 
in the FFLO state with $\qf \parallel \qia + \qib$. 
 We will discuss this point at the end of this paper.

\section{Neutron scattering}

 For a comparison with neutron scattering 
experiments~\cite{kenzelmann2008,kenzelmann2010}, 
we calculate the magnetic structure factor $|M(\q)|^{2}$, 
where $M(\q)$ is the Fourier transformed magnetic moment 
given as 
\begin{eqnarray}
&& M(\q) = \sum_{\i} M(\i) e^{{\rm i} \q \cdot \i}. 
\end{eqnarray}
 We here normalize $M(\i)$ so that $\sum_{\q} |M(\q)|^{2} =1$ to 
discuss the relative intensity of Bragg peaks. 
 Figures 3 and 4 show the magnetic structure factor in the AFM-FFLO 
states with $\qf \parallel \qia$ and $\qf \parallel \qia + \qib$, 
respectively. 
 The former corresponds to \Co under the magnetic field along 
$[1,\pm 1,0]$ direction, while the latter is realized 
in the magnetic field along $[1,0,0]$ or $[0,1,0]$ direction. 
 We have shown the figures rotated $45$ degree for a comparison 
with neutron scattering 
experiments~\cite{kenzelmann2008,kenzelmann2010}.

\begin{figure}[htbp]
\begin{center}
\includegraphics[width=15cm]{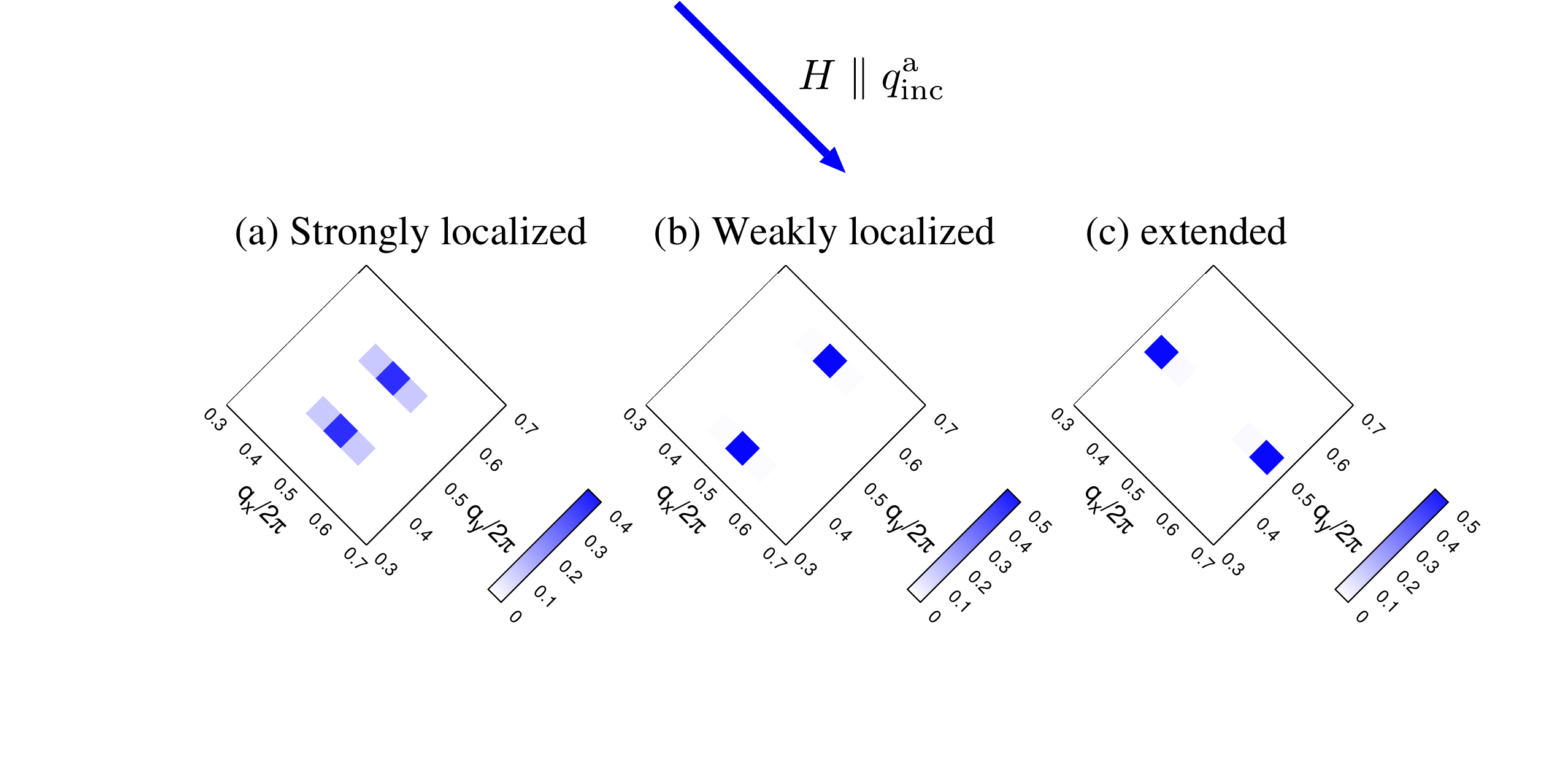}\hspace{1pc}%
\caption{
The magnetic structure factor $|M(\q)|^{2}$ in the AFM-FFLO state 
with $\qf \parallel \qia$. 
This AFM-FFLO state can be realized in \Co 
for the magnetic field along $[1,\pm 1,0]$ direction~\cite{kenzelmann2008}.  
We show the figures rotated by $45$ degrees for a comparison 
with neutron scattering experiments. 
The center of each figure shows $\q = \Q =(\pi,\pi)$, and therefore, 
the position from the center shows the incommensurate wave vector $\qi$. 
We assume the same parameters as in Fig.~1. 
}
\end{center}
\end{figure}
  
\begin{figure}[ht]
\begin{center}
\includegraphics[width=15cm]{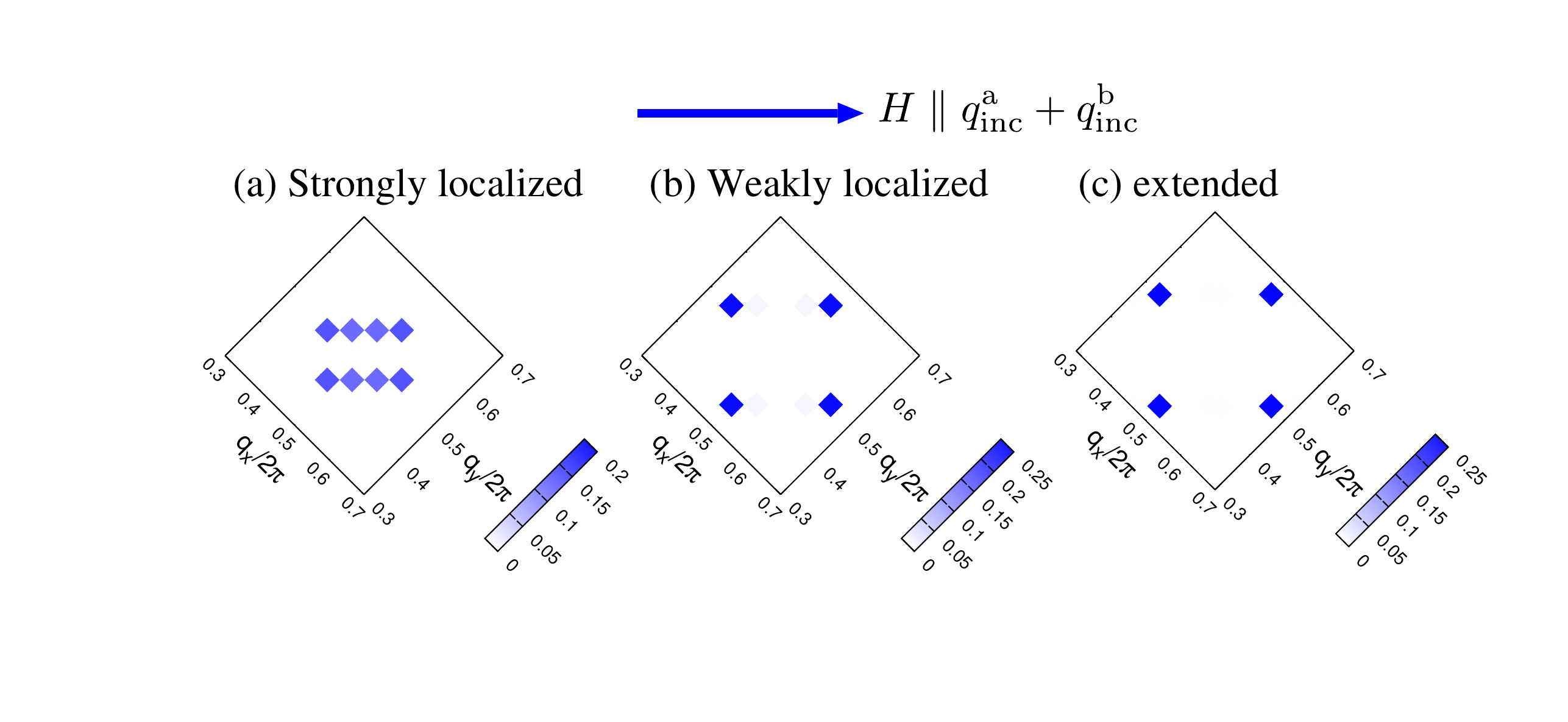}\hspace{1pc}%
\caption{
The magnetic structure factor $|M(\q)|^{2}$ in the AFM-FFLO state 
with $\qf \parallel \qia + \qib$. 
This AFM-FFLO state can be realized in \Co 
for the magnetic field along $[1,0,0]$ 
or $[0,1,0]$ direction~\cite{kenzelmann2010}.  
We assume the same parameters as in Fig.~3. 
}
\end{center}
\end{figure}

 In addition to the main Bragg peaks at $\q = \qia$, and/or $\q = \qib$, 
satellite peaks appear along the direction of 
magnetic field from the main peaks. 
 These satellite peaks arise from the broken translational symmetry 
in the FFLO state, and their amplitude reflects the spatial 
inhomogeneity of magnetic structure. 
 Therefore, the satellite peaks are pronounced in (I) the strongly 
localized case, while those are obscure in the cases (II) and (III). 
 
 We here discuss the possible magnetic structure of \Co in the HFSC phase 
on the basis of Figs.~3 and 4. 
 The strongly localized case (I) is incompatible with two features of 
neutron scattering experiments. 
First, the satellite peaks are absent or have a very weak intensity~\cite{kenzelmann2008,kenzelmann2010}. 
Second, the position of main Bragg peaks is 
independent of the orientation of magnetic field in the 
{\it ab}-plane~\cite{kenzelmann2010}. 
 The extended case (III) is incompatible with the neutron scattering 
measurement for magnetic fields along $[1,\pm 1,0]$ for which 
the incommensurate wave vector $\qi$ is perpendicular to the 
field $\vec{H}$~\cite{kenzelmann2008}. 
 On the other hand, the weakly localized case (II) shown in Figs.~3(b) 
and 4(b) is consistent with the neutron scattering experiments in which 
the incommensurate wave vector is perpendicular to the field direction 
$\qi \perp \vec{H}$ for $\vec{H} \parallel [1,\pm 1,0]$, and 
the position of main Bragg peaks does not change by rotating the 
magnetic field in the {\it ab}-plane. 
 The four main Bragg peaks appear in Fig.~4(b) in contrast 
to Fig.~3(b) owing to the symmetry of system. 
 This change has been observed in the neutron scattering 
measurement too~\cite{kenzelmann2010}. 
 According to these discussions, the magnetic structure in the 
possible AFM-FFLO state in \Co should be (II) the weakly localized 
case. This is the main conclusion of this paper.

\section{Discussion}

 We studied the magnetic structure of AFM order in the FFLO 
superconducting state. 
 We find that the spatial inhomogeneity of magnetic moment is 
reduced by increasing the incommensurability $|\qi|$ and enhancing 
the nesting of Fermi surface. This change of the magnetic structure 
is realized in our model by decreasing the number density from the half 
filling. 
 Comparing our results with the neutron scattering experiments in 
\Cof, the unconventional magnetic order in the HFSC phase of \Co is 
consistent with the AFM-FFLO state proposed by us when the 
AFM staggered moment is weakly localized around the FFLO nodal planes. 
 This is the ``weakly localized case'' in our classification of  the
magnetic structure.

 Finally, we discuss two points. 
(1) In our discussion we assumed that the FFLO modulation vector 
$\qf$ lies parallel to the magnetic field, since this is the most stable FFLO 
state, unless the anisotropy of Fermi surface favors another $\qf$.
On the other hand, if $\qf$ is fixed by the anisotropic electronic structure, 
the position of the Bragg peaks measured by neutron scattering
would be independent of the magnetic field orientation, consistent with the
experimental observation. This would apply to the AFM-FFLO state 
in the ``weakly localized case'' as well as in the ``strongly localized case'' 
and the ``extended case''. 
 This implies, however, that BCS-to-AFM-FFLO transition would be of first
order~\cite{adachi2003,ikeda:134504,ikeda:054517}. 
 While this is in contrast to the experimental observation, 
we can not exclude a weakly first order transition which is 
experimentally hard to detect.

(2) The magnetic structure for the magnetic 
field along $\qia + \qib$ corresponds to $\vec{H} \parallel [1,0,0]$ 
in \Cof. For this field direction Fig.~2 shows a double-$q$ 
Bragg peak structure of the magnetization 
$M_{\rm AF}(\i) \sim \cos(\qia \cdot \i) \pm \cos(\qib \cdot \i)$ 
which is different from the single-$q$ structure seen in Fig.~1. 
We distinguish two situations for the evolution of 
$M_{\rm AF}(\i) = \eta_a  \cos(\qia \cdot \i) + \eta_b \cos(\qib \cdot \i)$
for $\vec{H} \parallel \qia + \qib$. 
In the ''commensurate case'' where $\qia + \qib = 2 N \qf$, $ N $ 
being an integer, 
the double-$q$ structure ($ | \eta_a | = | \eta_b| $) remains stable 
for a small temperature range below $T_N $ 
and then changes continuously towards a single-$q$ structure, 
$M_{\rm AF}(\i) \sim \cos(\qia \cdot \i) $ or 
$M_{\rm AF}(\i) \sim \cos(\qib \cdot \i) $,  by shifting the weight between
the two components, $ | \eta_a | \neq | \eta_b| $. Note that this
corresponds to a symmetry breaking transition of $ Z_2 $ character. 
 On the other hand, for the ''incommensurate case'' 
$\qia + \qib \neq 2 N \qf $, we find $ \eta_a =0$ or $ \eta_b =0 $ 
(single-$q$ structure) for any 
$ T \leq T_N $. Since $ \qf $ depends
on the magnetic field, we expect to see a sequence of commensurate
points following the phase boundary $ T_N(H) $ in the $ H$-$T$ phase 
diagram. 
 This leads to a intriguing phase diagram near the commensurate points, 
which will be shown in another publication~\cite{yanase_unpublished}. 
 In most of the regime inside the AFM-FFLO phase the single-$q$ state is  
stable for magnetic fields $\vec{H} \parallel [1,0,0]$. 
 If this assumption holds true then the observed four Bragg peaks 
in neutron scattering should be interpreted as arising from domain formation of
the two degenerate single-$q$ states, i.e. $ |\eta_a| > | \eta_b| $ and $ |\eta_a| < | \eta_b| $. 
Note that the NMR spectrum in the HFSC phase of \Co \cite{curro2009} 
is consistent with a single-$q$ structure, but not with the double-$q$ structure. 
The detailed analysis of the NMR experiments under the assumption of weakly localized 
form of the AFM-FFLO state, as proposed in this paper, will be reported elsewhere.

\section*{Acknowledgements}
 The authors are grateful to D.F. Agterberg, R. Ikeda, M. Kenzelmann, K. Kumagai, 
K. Machida, Y. Matsuda,  V. F. Mitrovi\'c  and H. Tsunetsugu
for fruitful discussions. 
 This work was supported by 
a Grant-in-Aid for Scientific Research on Innovative Areas
``Heavy Electrons'' (No. 21102506) from MEXT, Japan. 
 It was also supported by a Grant-in-Aid for 
Young Scientists (B) (No. 20740187) from JSPS. 
 Numerical computation in this work was carried out 
at the Yukawa Institute Computer Facility. YY is grateful for the hospitality of the
Pauli Center of ETH Zurich. This work was also supported by the Swiss Nationalfonds
and the NCCR MaNEP.

\section*{References}
\bibliographystyle{unsrt}
\bibliography{FFLOAF}

\end{document}